\documentclass[twocolumn,showpacs,superscriptaddress,preprintnumbers,amsmath,amssymb]{revtex4-1}
\usepackage{graphicx}
\usepackage{dcolumn}
\usepackage{appendix}
\usepackage{bm}

\usepackage[usenames,dvipsnames]{xcolor}
\usepackage{subfigure}
\usepackage{bbm}
\usepackage{enumerate}
\usepackage[
  bookmarks=true,
  colorlinks,
  linkcolor=black,
  urlcolor=black,
  citecolor=black,
  plainpages=false,
  pdfpagelabels,
  final,
  breaklinks=true
]{hyperref}

\usepackage{titlesec}
\usepackage{array}
\usepackage{dsfont}
\usepackage{physics}

\begin{document}

\title{Absence of quantum optical coherence in high harmonic generation}
\author{Philipp Stammer}
\email{philipp.stammer@icfo.eu}
\affiliation{ICFO -- Institut de Ciencies Fotoniques, The Barcelona Institute of Science and Technology, 08860 Castelldefels (Barcelona), Spain}
\affiliation{Atominstitut, Technische Universität Wien, Stadionallee 2, 1020 Vienna, Austria}

\date{\today}

\begin{abstract}

The optical phase of the driving field in the process of high harmonic generation and the coherence properties of the harmonics are fundamental concepts in attosecond physics. 
Here, we consider to drive the process by incoherent classical and non-classical light fields exhibiting an undetermined optical phase. With this we introduce the notion of quantum optical coherence into high harmonic generation, and show that high harmonics can be generated from incoherent radiation despite having a vanishing electric field.
We explicitly derive the quantum state of the harmonics when driven by carrier-envelope phase unstable fields and show that the generated harmonics are incoherent and exhibiting zero electric field amplitudes. 
We find that the quantum state of each harmonic is diagonal in its photon number basis, but nevertheless has the exact same photon statistics as the widely considered coherent harmonics. From this we conclude that assuming coherent harmonic radiation can originate from a preferred ensemble fallacy. 
These findings have profound implications for attosecond experiments and how to infer about the harmonic radiation properties.

\end{abstract}

\maketitle

\textit{Introduction.}
High harmonic generation (HHG) is a parametric process in which an intense driving field is frequency up-converted with the resulting harmonic spectrum extending towards very high non-linear orders ranging from the infrared to the extreme-ultraviolet regime. 
In conventional HHG experiments the process is driven by a classical light source provided by a laser, while the description has almost exclusively focused on semi-classical approaches \cite{lewenstein1994theory}.
Furthermore, full quantum optical methods show that the generated harmonic radiation is coherent with the quantum state of the field modes given by product coherent states \cite{lewenstein2021generation, gorlach2020quantum, stammer2022high, rivera2022strong, stammer2023quantum}. This result holds in the limit of vanishing dipole moment correlations \cite{stammer2022high, stammer2022theory, stammer2023entanglement} and for the experimental boundary condition that the driving field is given by a pure initial coherent state. 
This assumption of an initial pure coherent state leads to a well defined phase in the associated classical driving field \cite{stammer2023iwo, stammer2023limitations} bridging the gap to the semi-classical picture \cite{stammer2023limitations}.
Closely related to the optical phase is the concept of optical coherence, which is associated to the statistical properties of the fluctuations of the light field \cite{mandel1965coherence, mandel1995optical}. 
Both of these concepts, the phase of the field and quantum optical coherence, will be scrutinized in this work for the process of HHG.
In particular, we focus on quantum optical coherence associated to the off-diagonal density matrix elements in the photon number basis of the corresponding field state. The discussion about the existence of optical coherence was initiated in \cite{molmer1997optical, m1997quantum}, with subsequent studies on the relevance of this optical coherence for quantum information processing protocols \cite{rudolph2001requirement}, and causes a debate about the proper description of the quantum state of a laser field \cite{molmer1997optical, rudolph2001requirement, bartlett2006dialogue, van2001quantum, van2001quantum2}.
The notion of quantum optical coherence is of particular importance for the rapidly growing interest in generating quantum light using HHG and its applications \cite{lewenstein2021generation, gorlach2020quantum, stammer2022high, stammer2023entanglement, pizzi2023light, gorlach2023high, martos2023metrological}. 

Approaches going beyond the semi-classical perspective for the description of HHG considered the quantum optical analog of driving the process with classical laser radiation given by coherent states \cite{gorlach2020quantum, lewenstein2021generation, rivera2022strong, stammer2022high, stammer2022theory, stammer2023quantum, gombkotHo2016quantum, gombkotHo2020high}, showing that the harmonic radiation is coherent as well. 
Even further, recent work on the quantum optical description of HHG studied the process when driving with non-classical states of light \cite{gorlach2023high, even2023photon}. 
For instance, light fields with a well defined photon number were considered, resulting in an arbitrary phase of the field.
Furthermore, this approach allows to consider light states with vanishing quantum optical coherence, i.e. a diagonal density matrix in the photon number basis, leading to a vanishing mean electric field value \cite{stammer2023limitations}.
Therefore, the analysis in the present work allows to pose questions such as: \textit{Can HHG be driven by light fields without quantum optical coherence, and if so, what are the coherence properties of the generated harmonics? For the experimental consequences, can we distinguish coherent and incoherent harmonic radiation from the measurement?}
In the following we will give definite answers to those questions, which have yet not been posed before. This is particularly important for virtually all attosecond experiments in which coherent harmonic radiation with oscillating electric field amplitudes is assumed. 
We discuss how this assumption can lead to a preferred ensemble fallacy in the interpretation of the measurement data, and we provide further insights into the radiation properties and the structure of the generated quantum state from HHG. 
Controlling the quantum state of the harmonic field modes is of current interest since the domain of strong field physics has recently become a tool for quantum state engineering \cite{lewenstein2021generation, stammer2023quantum, pizzi2023light} of high photon number entangled states \cite{stammer2022high, stammer2022theory} and coherent state superposition in terms of optical cat states with photon numbers sufficient to induce non-linear processes \cite{lamprou2023nonlinear}. Further, driving HHG in solid state \cite{rivera2022quantum, gonoskov2022nonclassical} or strongly correlated materials \cite{pizzi2023light} allows to obtain possibly interesting field states. 
Understanding the quantum coherence properties of the generated harmonic radiation, and to derive the associated quantum state is essential for connecting strong field physics with quantum optics and quantum information science \cite{lewenstein2022attosecond, bhattacharya2023strong, cruz2024quantum}.

\begin{figure}
    \centering
	\includegraphics[width=1\columnwidth]{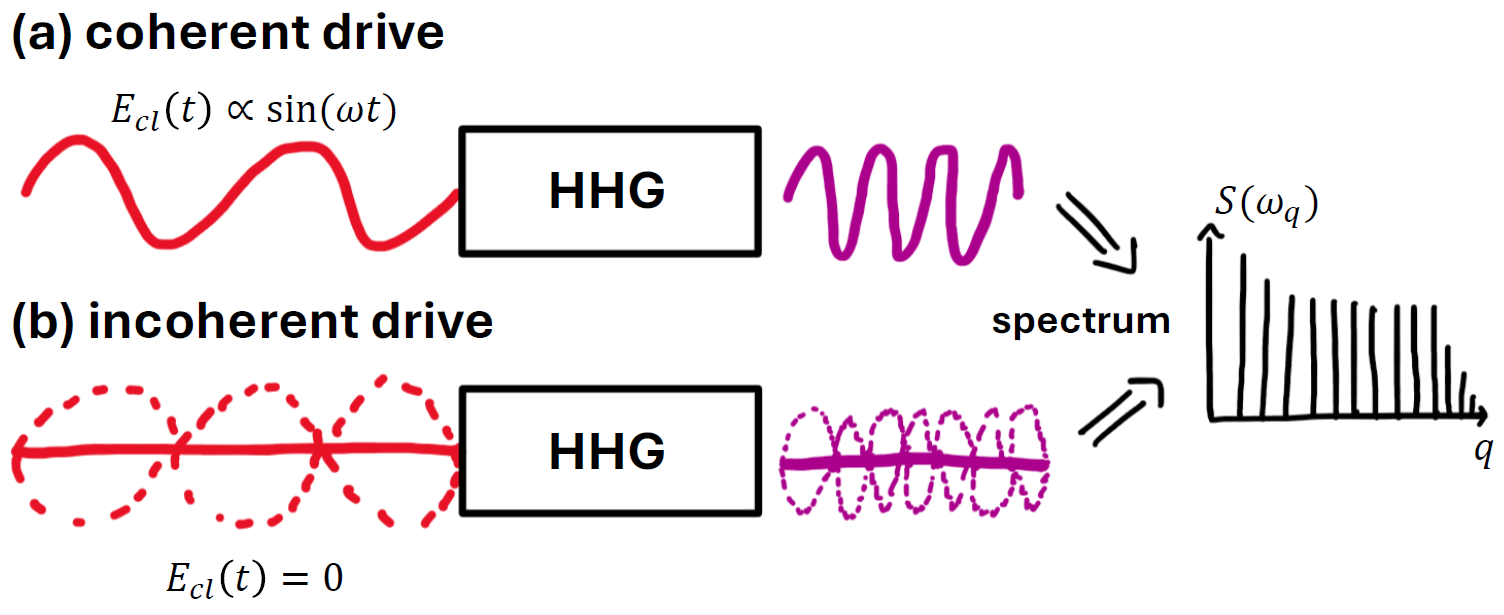}
	\caption{Graphical illustration of HHG driven by (a) a coherent field $\ket{\alpha}$, and (b) an incoherent field $\rho_{\abs{\alpha_0}}$. For the coherent driving field we have an oscillating classical electric field $E_{cl} (t) \propto \sin(\omega t)$, while for the incoherent fields the classical electric field vanishes $E_{cl} (t) = 0$. The resulting harmonic generation are coherent and incoherent, respectively, but lead to the identical HHG spectrum $S(\omega_q) \propto \expval*{a_q^\dagger a_q}$. }
      \label{fig:sketch}
\end{figure}

\textit{HHG driven by coherent light.}
Before analyzing the process of HHG driven by incoherent radiation, we first consider the case of driving the atom by classical coherent laser light. 
The quantum optical description of the experimental boundary condition of the coherent driving laser is given by an initial coherent state $\ket{\alpha}$, while the harmonic field modes $q$ are considered to be in the vacuum $\ket{\{ 0_q\} } = \otimes_q \ket{0_q}$. 
The coupling of the optical field modes to the electron is taken into account within the dipole approximation with the interaction Hamiltonian $H_I = - d E_Q(t)$, and electric field operator
\begin{align}
    E_Q(t) = - i \kappa \sum_q \sqrt{q} \left( a_q^\dagger e^{ i \omega_q t} - a_q e^{-i \omega_q t} \right),
\end{align}
where $\kappa \propto 1/\sqrt{V}$ is proportional to the quantization volume $V$. 
To solve the dynamics for the field modes a unitary transformation is performed \cite{lewenstein2021generation, stammer2023quantum}, which shifts the initial state of the driving field mode to the origin in phase space. This is done by using the displacement operator $D(\alpha)$ such that the interaction Hamiltonian obtains an additional term $H_{cl}(t) = - d E_{cl}(t)$, and the new initial state of the driving mode is given by the vacuum $D^\dagger (\alpha) \ket{\alpha} = \ket{0}$. This new term takes into account the fact that the initial driving laser mode is given by a coherent state and leads to the semi-classical interaction of the electron dipole moment with the classical electric field (see Fig.\ref{fig:sketch} (a) for an illustration of a classical field driving HHG)
\begin{align}
\label{eq:e_classical}
    E_{cl}(t) = \Tr[E_Q(t) \dyad{\alpha}] = i \kappa \left( \alpha e^{- i \omega t} - \alpha^* e^{i \omega t} \right),    
\end{align}
associated to the driving laser. 
This unitary transformation defines a semi-classical reference frame, which is unique for a pure coherent state initial condition since the phase $\phi = \operatorname{arg}(\alpha)$ of $\ket{\alpha}$ is well defined \cite{stammer2023iwo, stammer2023limitations}. 
Within this frame the dynamics of the optical field modes conditioned on HHG can be solved such that the evolution is given by a multi-mode displacement operation \cite{stammer2023quantum}. The final state of the harmonic field modes after the interaction is thus given by product coherent states
\begin{align}
\label{eq:linear_mapping}
    \ket{\{ 0_q \} } \to \prod_q D(\chi_q) \ket{ \{ 0_q \} } = \ket{ \{ \chi_q \} },  
\end{align}
with the amplitudes proportional to the Fourier transform (FT) of the time-dependent dipole moment expectation value in the ground state
\begin{align}
  \label{eq:amplitudes}
    \chi_q = - i \sqrt{q} \int dt \expval{d(t) } e^{i \omega_q t }, 
\end{align}
for the electron driven by the classical field \eqref{eq:e_classical}.
The fact that the final state is a pure state in terms of product coherent states comes from neglecting dipole moment correlations during the evolution \cite{sundaram1990high, stammer2022high, stammer2022theory}. This holds for small depletion of the electronic ground state, and it was shown that taking into account these dipole moment correlations leads to entanglement and squeezing of the optical field modes \cite{stammer2023entanglement}. 
We want to emphasize again that the linear mapping in \eqref{eq:linear_mapping} is based on the assumption of negligible dipole moment correlations, which was shown to be the relevant regime for HHG in atoms~\cite{sundaram1990high, lewenstein2021generation}. In contrast, driving correlated or solid state systems can result in correlations between the field modes~\cite{pizzi2023light, stammer2023entanglement}. Due to the high intensity of the driving field the induced charge current by means of the dipole moment expectation value $\expval{d(t)}$ is the dominant contribution to the emitted harmonic radiation, while higher order dipole moment transitions are much smaller. The non-linearity in the process of HHG lies within the highly non-linear oscillations of the charge current, and the FT of the dipole moment determines the harmonic amplitudes as seen from \eqref{eq:amplitudes}. 
In the following we discuss how the description changes when considering driving fields without a well defined phase such that the unitary transformation into the semi-classical frame is not uniquely defined anymore \cite{stammer2023limitations}.

\textit{Incoherent driving and the optical phase.}
To describe the process of HHG driven by incoherent light we shall first consider a classical light field by means of the mixture of coherent states over all phases 
\begin{align}
\label{eq:mixed1}
    \rho_{\abs{\alpha_0}} = \frac{1}{2\pi} \int_0^{2\pi} d\phi \dyad{\abs{\alpha_0} e^{i \phi}},
\end{align}
which in contrast to a pure coherent state $\ket{\alpha}$ has an arbitrary phase $\phi$. Due to the totally undetermined phase of the field, this state does not allow to uniquely define a semi-classical frame by means of the unitary displacement operation $D(\alpha)$. A consequence is that this field has a vanishing mean electric field value at all times (see Fig.\ref{fig:sketch}  (b) for a comparison with the coherent driving field)
\begin{align}
\label{eq:field_mixture}
    E_{cl}(t) & = \Tr[E_Q(t) \rho_{\abs{\alpha_0}}] = 0,
\end{align}
and the implications on the underlying semi-classical picture of HHG was discussed in \cite{stammer2023limitations}. However, despite the absence of a unique semi-classical frame one can express the initial state of the driving field in terms of phase-space distributions, which allows to decompose the field in terms of coherent states. Here, we shall focus on the generalized $P$-distribution $P(\alpha, \beta^*)$, allowing to write a quantum state in terms of a unique, positive and finite distribution function \cite{drummond1980generalised, kim1989properties, gilchrist1997positive}
\begin{align}
\label{eq:state_general}
    \rho = \int d^2\alpha d^2 \beta P(\alpha, \beta^*) \frac{\ket{\alpha} \bra{\beta}}{\bra{\beta} \ket{\alpha}}.
\end{align}

This allows to solve the HHG dynamics for an arbitrary initial light field \cite{gorlach2023high}, in close analogy to the approach used for a coherent state initial condition. The difference using the generalized $P$-representation is that there is not a single coherent state contribution, but due to the decomposition in \eqref{eq:state_general} each contribution of the coherent states $\ket{\alpha}$ and $\ket{\beta}$ driving the electron can be solved separately under the same approximations as in \cite{gorlach2020quantum, lewenstein2021generation, rivera2022strong}.
To derive the final field state generated from the electron currents driven by the distribution of intense fields, we use the general relation \cite{gilchrist1997positive, footnote6}
\begin{align}
\label{eq:positive_P}
    P(\alpha, \beta^*) = \frac{1}{4 \pi} e^{-\frac{\abs{\alpha - \beta^*}^2}{4}} Q\left( \frac{\alpha + \beta^*}{2} \right),
\end{align}
where $Q(\alpha) = \frac{1}{\pi} \bra{\alpha} \rho \ket{\alpha}$ is the Husimi $Q$ function of the driving field mode. Further, we take into account that the process is driven by light fields with sufficiently high intensities for generating harmonic radiation in a large enough quantization volume \cite{gorlach2023high, footnote2}. Hence, we consider the limit $\kappa \to 0$ and $\alpha \to \infty$ such that the physical electric field amplitude $\mathcal{E}_\alpha = 2\kappa \alpha$ remains finite, and evaluate the limits of the product in \eqref{eq:positive_P} separately 
\begin{align}
\label{eq:limit_exponential}
    \lim_{\kappa \to 0} \frac{1}{4\pi \kappa^2} e^{-\frac{\abs{\mathcal{E}_\alpha - \mathcal{E}_{\beta^*}}^2}{16 \kappa^2}} = \delta^{(2)}(\alpha - \beta^*).
\end{align}

Solving the dynamics of the electron currents and using the aforementioned limit, we find that the final field state after the end of the pulse is given by 
\begin{align}
\label{eq:solution}
    \rho = \int d^2 \alpha Q(\alpha) \prod_q \dyad{\chi_q(\alpha)}.
\end{align}

This final state describes an incoherent mixture of product coherent states over the driving field distribution given by $Q(\alpha)$ with product coherent states for each component of the driving field decomposition. 
The amplitudes are similarly as before
\begin{align}
\label{eq:chi_general}
    \chi_q(\alpha) = - i \sqrt{q} \int dt \expval{d_\alpha(t)} e^{i \omega_q t},
\end{align}
where $\expval{d_\alpha(t)}$ is the time-dependent dipole moment expectation value of the electron driven by the classical field of associated coherent state amplitude $\alpha$ from the decomposition of the initial driving field via $Q(\alpha)$. 
The coherent state amplitudes of the harmonic modes are the same as in the case of the pure coherent state driving field, just that the final state in \eqref{eq:solution} is incoherently mixed over the different coherent state contributions. 
With the final field state in \eqref{eq:solution} we can now compute the HHG spectrum $S(\omega_q) \propto \expval{a_q^\dagger a_q}$ for an arbitrary driving field
\begin{align}
   \expval{a_q^\dagger a_q} = \int d^2\alpha Q(\alpha) \abs{\chi_q(\alpha)}^2,
\end{align}
which is an incoherent average over the amplitudes $\abs{\chi_q(\alpha)}^2$ weighted by the Husimi distribution $Q(\alpha)$. 
Using that the Husimi distribution for the incoherent drive in \eqref{eq:mixed1} is given by 
\begin{align}
\label{eq:Q_mixture}
    Q_{\abs{\alpha_0}}(\alpha)  = \frac{1}{2\pi^2} \int_0^{2\pi} d\phi e^{-  \abs{\alpha - \alpha_0(\phi)}^2 },
\end{align}
we have 
\begin{align}
    \expval{a_q^\dagger a_q} = \frac{1}{2\pi^2} \int_0^{2\pi} d\phi \int d^2\alpha e^{-  \abs{\alpha - \alpha_0(\phi)}^2 } \abs{\chi_q(\alpha)}^2.
\end{align}

Since both $Q_{\abs{\alpha_0}}(\alpha) \ge 0$ and $\abs{\chi_q(\alpha)}^2 \ge 0$ for all $\alpha$ we find, despite the averaging over the phase $\phi$, that the spectrum is non-vanishing. This is particularly interesting because in contrast to the vanishing mean electric field value \eqref{eq:field_mixture} the spectrum does not vanish when averaging over all phases \cite{stammer2023limitations}.
This is the case because we incoherently average over the positive distribution $Q(\alpha)$ which does not allow for interference between the different contributions and thus there is no possible cancellation of different dipole currents of opposite phase. This is in fact a consequence of the limit performed in \eqref{eq:limit_exponential}, which holds for sufficiently intense fields and is necessary to drive the highly non-linear process of HHG. 

So far we have analyzed driving HHG by a classical field without optical coherence given by $\rho_{\abs{\alpha_0}}$. We shall now consider a genuinely non-classical field state without optical coherence by means of a photon number state $\ket{n}$ with sufficient intensity (limit of large $n$). 
Since \eqref{eq:solution} is the general solution for an arbitrary intense light field, we can use that the $Q$ function for the photon number state is given by 
\begin{align}
    Q_n(\alpha) = \frac{1}{\pi} \frac{\abs{\alpha}^{2n}}{n!} e^{-\abs{\alpha}^2} ,
\end{align}
such that the final state reads
\begin{align}
    \rho = \frac{1}{\pi} \int d^2\alpha \frac{\abs{\alpha}^{2n}}{n!} e^{-\abs{\alpha}^2} \prod_q \dyad{\chi_q(\alpha)}.
\end{align}

The HHG spectrum obtained from this state is proportional to 
\begin{align}
    \expval{a_q^\dagger a_q}_n = \frac{1}{\pi} \int d^2\alpha \frac{\abs{\alpha}^{2n}}{n!} e^{-\abs{\alpha}^2} \abs{\chi_q(\alpha)}^2,
\end{align}
which suggests that intense photon number states can drive the process of HHG \cite{gorlach2023high}. 
However, there is an interesting observation if one consistenly considers the limit used to obtain \eqref{eq:solution}, which is given by $\kappa \to 0$ for constant $\mathcal{E}_\alpha = 2 \kappa \alpha$. We can write the Husimi function $Q_n(\alpha)$ in terms of the field amplitude $\mathcal{E}_\alpha$ and take the respective limit such that
\begin{align}
    \lim_{\kappa \to 0} Q_n(\mathcal{E}_\alpha/(2 \kappa)) \frac{ d^2 \mathcal{E}_\alpha}{4\kappa^2} \propto \abs{\mathcal{E}_\alpha}^{2n} \delta^{(2)} (\mathcal{E_\alpha}) d^2 \mathcal{E}_\alpha, 
\end{align}
and consequently the HHG spectrum would read
\begin{align}
    \expval{a_q^\dagger a_q}_n  & \propto  \int d^2 \mathcal{E}_\alpha \abs{\mathcal{E}_\alpha}^{2n} \delta^{(2)} (\mathcal{E}_\alpha) \abs{\chi_q(\mathcal{E}_\alpha)}^2 \nonumber \\
     & \left. = \left[ \abs{\mathcal{E}_\alpha}^{2n} \abs{\chi_q(\mathcal{E}_\alpha)}^2 \right] \right|_{\mathcal{E}_\alpha = 0} .
\end{align}

This corresponds to the harmonic amplitudes $\chi_q(\mathcal{E}_\alpha)$ and the physical electric field amplitude $\mathcal{E}_\alpha$ evaluated at $\mathcal{E}_\alpha = 0$. However, already the harmonic amplitudes obtained from the semi-classical dipole moment expectation value in \eqref{eq:chi_general}, driven by the classical field $\mathcal{E}_\alpha =0$, would lead to a vanishing dipole moment, and thus, a vanishing harmonic spectrum. This implies that photon number states are not capable of driving the process of HHG in the limit used to obtain the general result \eqref{eq:solution}.

\textit{Quantum optical coherence in HHG.}
We have seen that driving the process of HHG with a mixture of coherent states over all phases $\rho_{\abs{\alpha_0}}$ is still possible despite the vanishing mean electric field amplitude. 
In the following we discuss an other crucial consequence of this observation. Interestingly, the mixed driving state in \eqref{eq:mixed1} does not exhibit quantum optical coherence in the sense of non-vanishing off-diagonal density matrix elements in the photon number basis. This can be seen when rewriting the mixture 
\begin{align}
    \rho_{\abs{\alpha_0}} = e^{- \abs{\alpha_0}^2} \sum_n \frac{\abs{\alpha_0}^{2n}}{n!} \dyad{n},
\end{align}
which is diagonal in the Fock basis and does therefore not have quantum optical coherence \cite{molmer1997optical, van2001quantum}. 
Since we have seen that this driving field state allows to generate high harmonic radiation for sufficiently large field intensities, it is now of interest to analyze the coherence properties of the harmonic radiation in the case of driving the process by light fields without optical coherence. 
This allows to answer the question: \textit{What are the quantum coherence properties of the harmonic radiation when driven by incoherent radiation?}

We therefore look at a single harmonic mode $q$ by tracing the state \eqref{eq:solution} over the remaining modes $q^\prime \neq q$. Since each state in the mixture is a product state we have
\begin{align}
    \rho_q = & \Tr_{q^\prime \neq q} [\rho] = \int d^2\alpha Q_{\abs{\alpha}}(\alpha) \dyad{\chi_q (\alpha)}.
\end{align}

We can now use the $Q$ function for the mixed initial state, and that in the limit of large field amplitudes $\mathcal{E}_\alpha$ considered above each exponential can be written as a $\delta$-function 
\begin{align}
    \lim_{\kappa \to 0} \frac{d^2 \mathcal{E}_\alpha}{4 \pi {\kappa}^2}e^{- \frac{\abs{\mathcal{E}_\alpha - \mathcal{E}_{\alpha_0}(\phi)}^2}{4 {\kappa}^2}} = \delta^{(2)}(\mathcal{E}_\alpha - \mathcal{E}_{\alpha_0}(\phi) ) d^2 \mathcal{E}_\alpha,   
\end{align}
such that we have 
\begin{align}
    \rho_q = \frac{1}{2\pi} \int_0^{2\pi} d\phi \dyad{ \chi_q(\mathcal{E}_{\alpha_0}(\phi))  }.
\end{align}

Expressing the state in the photon number basis we find
\begin{align}
    \rho_q = \frac{1}{2\pi^2} \int_0^{2\pi} d\phi e^{- \abs{\chi_q(\phi)}^2} \sum_{n,m} \frac{(\chi_q(\phi))^n (\chi_q^*(\phi))^m}{\sqrt{n! m!}} \dyad{n}{m},
\end{align}
where we have introduced the short hand notation $\chi_q(\phi) = \chi_q(\mathcal{E}_{\alpha_0}(\phi))$. To further simply the expression we use that for pulses of more than just a few cycles that the phase of the driving field, i.e. the carrier-envelope phase (CEP), only alters the phase of the induced dipole moment expectation value. Further, a different phase in the driving field can be seen as a time-delay $\Delta t = \phi/\omega$, such that for the harmonic amplitude we have
\begin{align}
    \chi_q(\phi) & = - i \sqrt{q} \int dt \expval{d_{|\alpha_0|}(t+\Delta t)} e^{i \omega_q t} \nonumber \\
    & = e^{- i \frac{\omega_q}{\omega} \phi} \chi_q (\abs{\alpha_0}).
\end{align}

And finally, the state of each harmonic field mode is given by 
\begin{align}
\label{eq:harmonic_q_final_incoherent}
    \rho_q = e^{- \abs{\chi_q(\abs{\alpha_0})}^2} \sum_{n} \frac{\abs{\chi_q(\abs{\alpha_0})}^{2n}}{n!} \dyad{n},
\end{align}
where we have used that 
\begin{align}
    \int_0^{2\pi} d\phi e^{- i \frac{\omega_q}{\omega}(n-m)\phi} = 2 \pi \delta (n-m).
\end{align}

We observe, that each harmonic field mode is diagonal in it's respective photon number basis and does not have quantum optical coherence by means of non-vanishing off-diagonal elements (the same would hold true for the case of an incoherent Fock state drive \cite{footnote4}). 
The observation that optical coherence, and non-vanishing electric field amplitudes, are not required to drive the process of high harmonic generation provides interesting insights into the underlying mechanism. This is because the harmonic field modes are still given by coherent states, which are generated by classical charge currents emitting coherent radiation \cite{footnote3}. In the case of HHG it is the electron current driven by the intense field which generates the coherent radiation. However, due to the incoherent averaging over all phases of the driving field, and consequently over all phases of the induced charge current, the final state of the harmonic field modes is incoherent, i.e. diagonal in the respective photon number basis. 
We emphasize that this incoherent state of each harmonic field mode originates despite the fact that the final state of all modes is in a product state, see Eq.\eqref{eq:solution}, and the mixture does not arise from a partial trace over an entangled state of all modes. 
However, we note that the final field state can be entangled when taking into account dipole moment correlations \cite{stammer2023entanglement}, which would also lead to mixed final states for each mode. Nevertheless, the effect considered here solely originates from the properties of the driving field and the role of the optical phase and coherence as discussed above. 

\textit{Optical coherence and the HHG spectrum.}
We now use this result to explicitly show that concluding on the coherence properties of the harmonic radiation in all experiments with CEP unstable driving fields and which solely measure the HHG spectrum are fallacious. This is particularly important because in virtually all descriptions of HHG experiments the generated harmonic radiation is assumed to be coherent, although the measurement of the spectrum alone does not allow to infer on the coherence properties of the generated harmonics. Thus, the commonly used assumption is no justified in these cases.
This is because the observer perspective of the spectrum does not distinguish between the coherent and incoherent harmonic radiation, which is because intensity measurements are only sensitive to the diagonal elements of the quantum state.
Therefore, the incoherent distribution in \eqref{eq:harmonic_q_final_incoherent} and a pure coherent state with the same amplitude give rise to the same spectrum. 
In more detail, this can be seen when computing the average photon number for the harmonic field mode in a pure coherent state $\bra{\chi_q}a_q^\dagger a_q\ket{\chi_q} = \abs{\chi_q}^2$, in comparison to the average for the incoherent state \eqref{eq:harmonic_q_final_incoherent} given by 
\begin{align}
    \expval{a_q^\dagger a_q} = \operatorname{Tr}[a_q^\dagger a_q \rho_q] = \abs{\chi_q}^2.
\end{align}

From the observation that the pure coherent state $\ket{\chi_q}$ and the incoherent state $\rho_q$ have the same mean photon number (and the same photon number distribution), we can conclude that the most used observable in HHG experiments, i.e. the HHG spectrum, is insensitive to the quantum optical coherence in the radiation field. 
While it is true that an intensity measurement is insensitive to optical coherence for any field state, we have shown here that in HHG the coherent and incoherent case have the same statistics and explicitly derived the incoherent state. 
Therefore, most HHG experiments can not distinguish between these two states.
This implies that inferring about the coherence properties of the harmonic radiation from the HHG spectrum alone, and using a coherent state description, can be fallacious by assuming a preferred ensemble in the description of the field state itself \cite{bartlett2006dialogue}.
This is particularly interesting when considering that the proper way of describing a CEP unstable driving field is rather given by the mixture $\rho_{\abs{\alpha_0}}$ than the pure state $\ket{\alpha}$ with a well defined phase.
A consequence of this is that interpreting the observation of the HHG spectrum by means of incoherent radiation is equally correct as using coherent radiation. This is because the process of HHG and detection of the spectrum alone is insensitive to quantum optical coherence.
Extending this analysis to other processes in attosecond science \cite{agostini2004physics} or non-linear optics \cite{kopylov2019study, li2023experimental}, such as harmonic generation driven by non-classical light \cite{spasibko2017multiphoton, kopylov2020spectral}, in which the field properties are discussed can lead to a new examination of those properties and its interpretation. 

\textit{Conclusions.}
The insights obtained when driving the process of HHG by incoherent radiation shows that quantum optical coherence, in terms of non-vanishing off-diagonal density matrix elements in the photon number basis, is not required to generate high-order harmonics.
However, the considerable difference to a coherent drive is that the emitted harmonic radiation is incoherent as well.
One reason why optical coherence is not required to drive HHG is because the different contributions of the driving field, by means of the distribution over coherent states, couple diagonally (incoherently) to the charge which emitts the harmonic radiation. This can be seen from \eqref{eq:solution} where the distribution of the incoherent average is given by the Husimi $Q$ function and performed over the coherent states into which the driving field is decomposed. 
This holds in the limit of intense fields with large amplitudes necessary for driving HHG. The process of HHG is only coherent by means of the emitted radiation due to the oscillating charge current of the electron for a driving field with a well defined phase. Averaging over all phases leads to vanishing quantum optical coherence. 
This suggests that further investigation about the role of the optical phase from a quantum optical perspective can provide insights into the properties of the generated harmonic radiation and many other processes in attosecond science. 
In particular, the role of the carrier-envelope phase (CEP) for ultrashort few-cycles pulses is of interest. Furthermore, this work highlights the importance of answering: \textit{what is the proper description of the experimental boundary condition, i.e. the quantum state, of an ultrashort few-cycle (CEP-stable) intense laser pulse?}
Moreover, this work shows that concluding on the coherence properties of the harmonic radiation from the observation of the spectrum alone, is not possible without falling into a preferred ensemble fallacy.
This is because the coherent and incoherent harmonic radiation exhibit the same photon statistics and most HHG experiments can not distinguish between these two.
Finally, we emphasize that it is not only a fallacy to conclude on the coherence properties of the harmonic radiation from the spectrum, but also to conclude on the mean field amplitude. The analysis in this work illustrates that harmonic radiation does not necessarily possess an electric field amplitude, and thus challenges the common believes about the radiation properties of high harmonic generation in attosecond experiments.
Therefore, we emphasize again, that the properties such as optical coherence in this case, depends on the observer perspective, i.e. the specific experiment to be performed. The observer perspective should be the first thing to be defined before talking about the properties of interest.

\begin{acknowledgments}

P.S. acknowledges funding from the European Union’s Horizon 2020 research and innovation programme under the Marie Skłodowska-Curie grant agreement No 847517. 
ICFO group acknowledges support from: ERC AdG NOQIA; MICIN/AEI (PGC2018-0910.13039/501100011033, CEX2019-000910-S/10.13039/501100011033, Plan National FIDEUA PID2019-106901GB-I00, FPI; MICIIN with funding from European Union NextGenerationEU (PRTR-C17.I1): QUANTERA MAQS PCI2019-111828-2); MCIN/AEI/ 10.13039/501100011033 and by the “European Union NextGeneration EU/PRTR" QUANTERA DYNAMITE PCI2022-132919 within the QuantERA II Programme that has received funding from the European Union’s Horizon 2020 research and innovation programme under Grant Agreement No 101017733Proyectos de I+D+I “Retos Colaboración” QUSPIN RTC2019-007196-7); Fundació Cellex; Fundació Mir-Puig; Generalitat de Catalunya (European Social Fund FEDER and CERCA program, AGAUR Grant No. 2021 SGR 01452, QuantumCAT \ U16-011424, co-funded by ERDF Operational Program of Catalonia 2014-2020); Barcelona Supercomputing Center MareNostrum (FI-2023-1-0013); EU (PASQuanS2.1, 101113690); EU Horizon 2020 FET-OPEN OPTOlogic (Grant No 899794); EU Horizon Europe Program (Grant Agreement 101080086 — NeQST), National Science Centre, Poland (Symfonia Grant No. 2016/20/W/ST4/00314); ICFO Internal “QuantumGaudi” project; European Union’s Horizon 2020 research and innovation program under the Marie-Skłodowska-Curie grant agreement No 101029393 (STREDCH) and No 847648 (“La Caixa” Junior Leaders fellowships ID100010434: LCF/BQ/PI19/11690013, LCF/BQ/PI20/11760031, LCF/BQ/PR20/11770012, LCF/BQ/PR21/11840013). Views and opinions expressed are, however, those of the author(s) only and do not necessarily reflect those of the European Union, European Commission, European Climate, Infrastructure and Environment Executive Agency (CINEA), nor any other granting authority. Neither the European Union nor any granting authority can be held responsible for them. 

\end{acknowledgments}

\bibliographystyle{unsrt}
\bibliography{literatur}{}

\begin{thebibliography}{10}

\bibitem{lewenstein1994theory}
Maciej Lewenstein, Ph~Balcou, M~Yu Ivanov, Anne L’huillier, and Paul~B
  Corkum.
\newblock Theory of high-harmonic generation by low-frequency laser fields.
\newblock {\em Physical Review A}, 49(3):2117, 1994.

\bibitem{lewenstein2021generation}
Maciej Lewenstein, Marcelo~F Ciappina, Emilio Pisanty, Javier Rivera-Dean,
  Philipp Stammer, Th~Lamprou, and Paraskevas Tzallas.
\newblock Generation of optical schr{\"o}dinger cat states in intense
  laser--matter interactions.
\newblock {\em Nature Physics}, 17(10):1104--1108, 2021.

\bibitem{gorlach2020quantum}
Alexey Gorlach, Ofer Neufeld, Nicholas Rivera, Oren Cohen, and Ido Kaminer.
\newblock The quantum-optical nature of high harmonic generation.
\newblock {\em Nature communications}, 11(1):4598, 2020.

\bibitem{stammer2022high}
Philipp Stammer, Javier Rivera-Dean, Theocharis Lamprou, Emilio Pisanty,
  Marcelo~F Ciappina, Paraskevas Tzallas, and Maciej Lewenstein.
\newblock High photon number entangled states and coherent state superposition
  from the extreme ultraviolet to the far infrared.
\newblock {\em Physical Review Letters}, 128(12):123603, 2022.

\bibitem{rivera2022strong}
Javier Rivera-Dean, Th~Lamprou, Emilio Pisanty, Philipp Stammer, Andr{\'e}s~F
  Ord{\'o}{\~n}ez, AS~Maxwell, MF~Ciappina, M~Lewenstein, and P~Tzallas.
\newblock Strong laser fields and their power to generate controllable
  high-photon-number coherent-state superpositions.
\newblock {\em Physical Review A}, 105(3):033714, 2022.

\bibitem{stammer2023quantum}
Philipp Stammer, Javier Rivera-Dean, Andrew Maxwell, Theocharis Lamprou,
  Andr{\'e}s Ord{\'o}{\~n}ez, Marcelo~F Ciappina, Paraskevas Tzallas, and
  Maciej Lewenstein.
\newblock Quantum electrodynamics of intense laser-matter interactions: a tool
  for quantum state engineering.
\newblock {\em PRX Quantum}, 4(1):010201, 2023.

\bibitem{stammer2022theory}
Philipp Stammer.
\newblock Theory of entanglement and measurement in high-order harmonic
  generation.
\newblock {\em Physical Review A}, 106(5):L050402, 2022.

\bibitem{stammer2023entanglement}
Philipp Stammer, Javier Rivera-Dean, Andrew~S Maxwell, Theocharis Lamprou,
  Javier Arg{\"u}ello-Luengo, Paraskevas Tzallas, Marcelo~F Ciappina, and
  Maciej Lewenstein.
\newblock Entanglement and squeezing of the optical field modes in high
  harmonic generation.
\newblock {\em Physical Review Letters}, 132(14):143603, 2024.

\bibitem{stammer2023iwo}
P~Stammer and M~Lewenstein.
\newblock Quantum optics as applied quantum electrodynamics is back in town.
\newblock {\em Acta Physica Polonica A ISSN 1898-794X}, 143(6):S42--S42, 2023.

\bibitem{stammer2023limitations}
Philipp Stammer.
\newblock On the limitations of the semi-classical picture in high harmonic
  generation.
\newblock {\em arXiv:2308.15087}, 2023.

\bibitem{mandel1965coherence}
Leonard Mandel and Emil Wolf.
\newblock Coherence properties of optical fields.
\newblock {\em Reviews of modern physics}, 37(2):231, 1965.

\bibitem{mandel1995optical}
Leonard Mandel and Emil Wolf.
\newblock {\em Optical coherence and quantum optics}.
\newblock Cambridge university press, 1995.

\bibitem{molmer1997optical}
Klaus M{\o}lmer.
\newblock Optical coherence: A convenient fiction.
\newblock {\em Physical Review A}, 55(4):3195, 1997.

\bibitem{m1997quantum}
Klaus M{\o}lmer.
\newblock Quantum entanglement and classical behaviour.
\newblock {\em Journal of Modern Optics}, 44(10):1937--1956, 1997.

\bibitem{rudolph2001requirement}
Terry Rudolph and Barry~C Sanders.
\newblock Requirement of optical coherence for continuous-variable quantum
  teleportation.
\newblock {\em Physical Review Letters}, 87(7):077903, 2001.

\bibitem{bartlett2006dialogue}
Stephen~D Bartlett, Terry Rudolph, and Robert~W Spekkens.
\newblock Dialogue concerning two views on quantum coherence: factist and
  fictionist.
\newblock {\em International Journal of Quantum Information}, 4(01):17--43,
  2006.

\bibitem{van2001quantum}
SJ~Van~Enk and Christopher~A Fuchs.
\newblock Quantum state of an ideal propagating laser field.
\newblock {\em Physical review letters}, 88(2):027902, 2001.

\bibitem{van2001quantum2}
Steven~J van Enk and Christopher~A Fuchs.
\newblock The quantum state of a propagating laser field.
\newblock {\em quant-ph/0111157}, 2001.

\bibitem{pizzi2023light}
Andrea Pizzi, Alexey Gorlach, Nicholas Rivera, Andreas Nunnenkamp, and Ido
  Kaminer.
\newblock Light emission from strongly driven many-body systems.
\newblock {\em Nature Physics}, 19(4):551--561, 2023.

\bibitem{gorlach2023high}
Alexey Gorlach, Matan~Even Tzur, Michael Birk, Michael Kr{\"u}ger, Nicholas
  Rivera, Oren Cohen, and Ido Kaminer.
\newblock High-harmonic generation driven by quantum light.
\newblock {\em Nature Physics}, pages 1--8, 2023.

\bibitem{martos2023metrological}
Tom{\'a}s~Fern{\'a}ndez Martos, Maciej Lewenstein, Grzegorz
  Rajchel-Mieldzio{\'c}, and Philipp Stammer.
\newblock Metrological robustness of high photon number optical cat states.
\newblock {\em arXiv:2311.01371}, 2023.

\bibitem{gombkotHo2016quantum}
{\'A}kos Gombk{\"o}t{\H{o}}, Attila Czirj{\'a}k, S{\'a}ndor Varr{\'o}, and
  P{\'e}ter F{\"o}ldi.
\newblock Quantum-optical model for the dynamics of high-order-harmonic
  generation.
\newblock {\em Physical Review A}, 94(1):013853, 2016.

\bibitem{gombkotHo2020high}
{\'A}kos Gombk{\"o}t{\H{o}}, S{\'a}ndor Varr{\'o}, P{\'e}ter Mati, and
  P{\'e}ter F{\"o}ldi.
\newblock High-order harmonic generation as induced by a quantized field:
  Phase-space picture.
\newblock {\em Physical Review A}, 101(1):013418, 2020.

\bibitem{even2023photon}
Matan Even~Tzur, Michael Birk, Alexey Gorlach, Michael Kr{\"u}ger, Ido Kaminer,
  and Oren Cohen.
\newblock Photon-statistics force in ultrafast electron dynamics.
\newblock {\em Nature Photonics}, 17(6):501--509, 2023.

\bibitem{lamprou2023nonlinear}
Theocharis Lamprou, Javier Rivera-Dean, Philipp Stammer, Maciej Lewenstein, and
  Paraskevas Tzallas.
\newblock Nonlinear optics using intense optical schr$\backslash$" odinger"
  cat" states.
\newblock {\em arXiv:2306.14480}, 2023.

\bibitem{rivera2022quantum}
Javier Rivera-Dean, Philipp Stammer, Andrew~S Maxwell, Theocharis Lamprou,
  Andr{\'e}s~F Ord{\'o}{\~n}ez, Emilio Pisanty, Paraskevas Tzallas, Maciej
  Lewenstein, and Marcelo~F Ciappina.
\newblock Quantum optical analysis of high-harmonic generation in solids within
  a wannier-bloch picture.
\newblock {\em arXiv:2211.00033}, 2022.

\bibitem{gonoskov2022nonclassical}
Ivan Gonoskov, Ren{\'e} Sondenheimer, Christian H{\"u}necke, Daniil Kartashov,
  Ulf Peschel, and Stefanie Gr{\"a}fe.
\newblock Nonclassical light generation and control from laser-driven
  semiconductor intraband excitations.
\newblock {\em Physical Review B}, 109(12):125110, 2024.

\bibitem{lewenstein2022attosecond}
Maciej Lewenstein, Niccolo Baldelli, Utso Bhattacharya, Jens Biegert,
  Marcelo~Fabi{\'a}n Ciappina, T~Grass, Piotr~Tadeusz Grochowski, AS~Johnson,
  Th~Lamprou, Andrew~S Maxwell, et~al.
\newblock Attosecond physics and quantum information science.
\newblock In {\em International Conference on Attosecond Science and
  Technology}, pages 27--44. Springer, 2012.

\bibitem{bhattacharya2023strong}
Utso Bhattacharya, Theocharis Lamprou, Andrew~Stephen Maxwell, Andres Ordonez,
  Emilio Pisanty, Javier Rivera-Dean, Philipp Stammer, Marcelo~F Ciappina,
  Maciej Lewenstein, and Paraskevas Tzallas.
\newblock Strong--laser--field physics, non--classical light states and quantum
  information science.
\newblock {\em Reports on Progress in Physics}, 2023.

\bibitem{cruz2024quantum}
Lidice Cruz-Rodriguez, Diptesh Dey, Antonia Freibert, and Philipp Stammer.
\newblock Quantum phenomena in attosecond science.
\newblock {\em arXiv:2403.05482}, 2024.

\bibitem{sundaram1990high}
Bala Sundaram and Peter~W Milonni.
\newblock High-order harmonic generation: simplified model and relevance of
  single-atom theories to experiment.
\newblock {\em Physical Review A}, 41(11):6571, 1990.

\bibitem{drummond1980generalised}
PD~Drummond and CW~Gardiner.
\newblock Generalised p-representations in quantum optics.
\newblock {\em Journal of Physics A: Mathematical and General}, 13(7):2353,
  1980.

\bibitem{kim1989properties}
MS~Kim, FAM De~Oliveira, and PL~Knight.
\newblock Properties of squeezed number states and squeezed thermal states.
\newblock {\em Physical Review A}, 40(5):2494, 1989.

\bibitem{gilchrist1997positive}
A~Gilchrist, CW~Gardiner, and PD~Drummond.
\newblock Positive p representation: Application and validity.
\newblock {\em Physical Review A}, 55(4):3014, 1997.

\bibitem{footnote6}
This relation holds if the Fourier coefficients of $P(\alpha, \beta^*)$ vanish
  faster than any power of $\abs{\alpha}$ and $\abs{\beta}$, which is the case
  for the states under consideration in this work \cite{olsen2009numerical}.

\bibitem{footnote2}
We consider the limit of high intense laser fields in which the physical
  quantity of the electric field amplitude $\mathcal{E}_\alpha$ remains fiinte
  and constant in the limit of large quantization volumes $\kappa \to 0$ and
  large coherent state amplitudes $\alpha \to \infty $.

\bibitem{footnote4}
For the Fock state drive in the considered limit we have $\rho_q =
  \dyad{\chi_q(0)} = \dyad{0}$, which is evidently diagonal in the photon
  number basis.

\bibitem{footnote3}
The first mentioning (to the best of our knowledge) of the quantum optical
  coherence properties in HHG due to the associated classical charge current is
  given in \cite{molmer1997optical}.

\bibitem{agostini2004physics}
Pierre Agostini and Louis~F DiMauro.
\newblock The physics of attosecond light pulses.
\newblock {\em Reports on progress in physics}, 67(6):813, 2004.

\bibitem{kopylov2019study}
Denis~A Kopylov, Kirill~Yu Spasibko, Tatiana~V Murzina, and Maria~V Chekhova.
\newblock Study of broadband multimode light via non-phase-matched sum
  frequency generation.
\newblock {\em New Journal of Physics}, 21(3):033024, 2019.

\bibitem{li2023experimental}
Cheng Li, Boris Braverman, Girish Kulkarni, and Robert~W Boyd.
\newblock Experimental generation of polarization entanglement from spontaneous
  parametric down-conversion pumped by spatiotemporally highly incoherent
  light.
\newblock {\em Physical Review A}, 107(4):L041701, 2023.

\bibitem{spasibko2017multiphoton}
Kirill~Yu Spasibko, Denis~A Kopylov, Victor~L Krutyanskiy, Tatiana~V Murzina,
  Gerd Leuchs, and Maria~V Chekhova.
\newblock Multiphoton effects enhanced due to ultrafast photon-number
  fluctuations.
\newblock {\em Physical Review Letters}, 119(22):223603, 2017.

\bibitem{kopylov2020spectral}
Denis~A Kopylov, Andrei~V Rasputnyi, Tatiana~V Murzina, and Maria~V Chekhova.
\newblock Spectral properties of second, third and fourth harmonics generation
  from broadband multimode bright squeezed vacuum.
\newblock {\em Laser Physics Letters}, 17(7):075401, 2020.

\bibitem{olsen2009numerical}
MK~Olsen and AS~Bradley.
\newblock Numerical representation of quantum states in the positive-p and
  wigner representations.
\newblock {\em Optics Communications}, 282(19):3924--3929, 2009.

\end{thebibliography}


\end{document}